\documentclass[aps, superscriptaddress, showpacs, twocolumn]{revtex4-1}
\usepackage{ORI_Group_style}

\hypersetup{colorlinks,linkcolor=blue,citecolor=blue,urlcolor=blue}

\begin{document}
\title{Generation of Spin-Wave Pulses by Inverse Design}
\author{S. Casulleras}
\email{silvia.casulleras-guardia@uibk.ac.at}
\affiliation{Institute for Quantum Optics and Quantum Information of the Austrian Academy of Sciences, 6020 Innsbruck, Austria}
\affiliation{Institute for Theoretical Physics, University of Innsbruck, 6020 Innsbruck, Austria}

\author{S. Knauer}
\author{Q. Wang}
\affiliation{Faculty of Physics, University of Vienna, 1090 Vienna, Austria}

\author{O. Romero-Isart}
\affiliation{Institute for Quantum Optics and Quantum Information of the Austrian Academy of Sciences, 6020 Innsbruck, Austria}
\affiliation{Institute for Theoretical Physics, University of Innsbruck, 6020 Innsbruck, Austria}

\author{A.V. Chumak}
\affiliation{Faculty of Physics, University of Vienna, 1090 Vienna, Austria}

\author{C. Gonzalez-Ballestero}
\email{carlos.gonzalez-ballestero@uibk.ac.at}
\affiliation{Institute for Quantum Optics and Quantum Information of the Austrian Academy of Sciences, 6020 Innsbruck, Austria}
\affiliation{Institute for Theoretical Physics, University of Innsbruck, 6020 Innsbruck, Austria}

\begin{abstract}
    The development of fast magnonic information processing nanodevices requires operating with short spin-wave pulses, but, the shorter the pulses, the more affected they are by information loss due to broadening and dispersion. The capability of engineering spin-wave pulses and controlling their propagation could solve this problem. Here, we provide a method to generate linear spin-wave pulses with a desired spatial-temporal profile in magnonic waveguides based on inverse design. As relevant examples, we theoretically predict that both rectangular and self-compressing spin-wave pulses can be generated in state-of-the-art waveguides with fidelities $\gtrsim96\%$ using narrow stripline antennas. The method requires minimal computational overhead and is universal, i.e., it applies to arbitrary targeted pulse shapes, type of waves (exchange or dipolar), waveguide materials, and waveguide geometries. It can also be extended to more complex magnonic structures. Our results could lead to the utilization of large-scale magnonic circuits for classical and quantum information processing.
\end{abstract}

\maketitle

\section*{Introduction}

Spin waves, exhibiting strong non-linearity and low loss coefficients, are promising candidates for classical and quantum
information processing~\cite{Barman_roadmap,9706176,Lachance_Quirion_2019} and  surpass their  electric current-based counterpart by harnessing properties such as frequency and phase~\cite{10.1038/nphys3347,10.1038/s41578-021-00332-w,Rana2019a, Chen_2021,YU20211}.
The demonstrations of coherent spin-wave transport in nanoscale magnetic structures and prototype devices~\cite{10.1021/acs.nanolett.0c00657,PhysRevApplied.16.024028,10.1038/s42005-018-0056-x,doi:10.1126/sciadv.abb4042, 10.1038/s41928-020-00485-6} allow access to the further miniaturization of large-scale magnonic circuits. Most of these works focus on tailoring spin-wave propagation through spatial-temporal nanostructure engineering~\cite{10.1038/nphys3325,10.1038/s41467-021-22897-4,PhysRevApplied.16.024028,10.1038/s42005-018-0056-x}  or by coupling to other systems such as paramagnetic spins~\cite{PhysRevB.105.075410, PRXQuantum.2.040314,doi:10.1126/sciadv.abd3556,acs.nanolett.1c02654} or acoustic waves~\cite{Kryshtal_2017,PhysRevB.81.140404}. Recently, the concept of inverse design has been introduced into magnonics numerically~\cite{10.1038/s41467-021-22897-4,Papp2021} and experimentally~\cite{Kiechle_22}, and has shown its great potential for radio frequency applications as well as for Boolean and neuromorphic computing. In these investigations, a medium through which a spin wave propagates was designed  to obtain the required functionality, while the wave itself was excited continuously. However, modern high-performance computing demands the use of short pulses that carry data at high clock rates. The fact that spin-wave dispersions are not linear and the different spectral components of the pulse have different group velocities leads to a broadening and distortion of the pulse shape, resulting in data loss. The minimum duration of the pulse is also limited by the bandwidth of the spin-wave spectrum.  

Here we propose an inverse design method (IDM) to obtain linear spin-wave pulses of arbitrary target shape at any point of a waveguide. The method is universal, i.e., it is suitable for both dipolar or exchange spin waves and for waveguides made of any material and with any geometry. As examples, we theoretically demonstrate the generation of self-compressing and rectangular pulses in state-of-the-art yttrium-iron-garnet (YIG) nanowaveguides. The self-compressing pulses possess an increased amplitude in a defined local point of a waveguide, enabling addressed read/write of data or local triggering of nonlinear phenomena in classical and quantum magnonic networks. The rectangular-wave pulses are of great interest for radio frequency and binary data processing as they allow the highest (undisturbed) data transmission rate.  It is shown that the IDM provides a pulse generation fidelity close to unity. 

This article is organized as follows. First, we provide a stepwise method for the determination of the voltage signal which must be applied to a narrow but arbitrarily-shaped antenna to generate an arbitrary target spin-wave pulse, using minimal micromagnetic simulations. Then, we illustrate our method for three particular examples, namely the generation of a self-compressing spin-wave pulse and a rectangular pulse, both in the exchange regime of a recently reported YIG nanowaveguide~\cite{doi:10.1063/5.0045570}, and of a weakly self-compressing spin-wave pulse in the dipolar regime. Our results are verified by micromagnetic simulations using MuMax3~\cite{mumax3}. Additionally, 
we discuss both the pulse generation fidelity and the energy cost for pulse generation as a function of antenna size. Finally, a discussion of our results is presented in the Conclusion section.

\begin{figure}[t]
\noindent \begin{centering}
\includegraphics[scale=1]{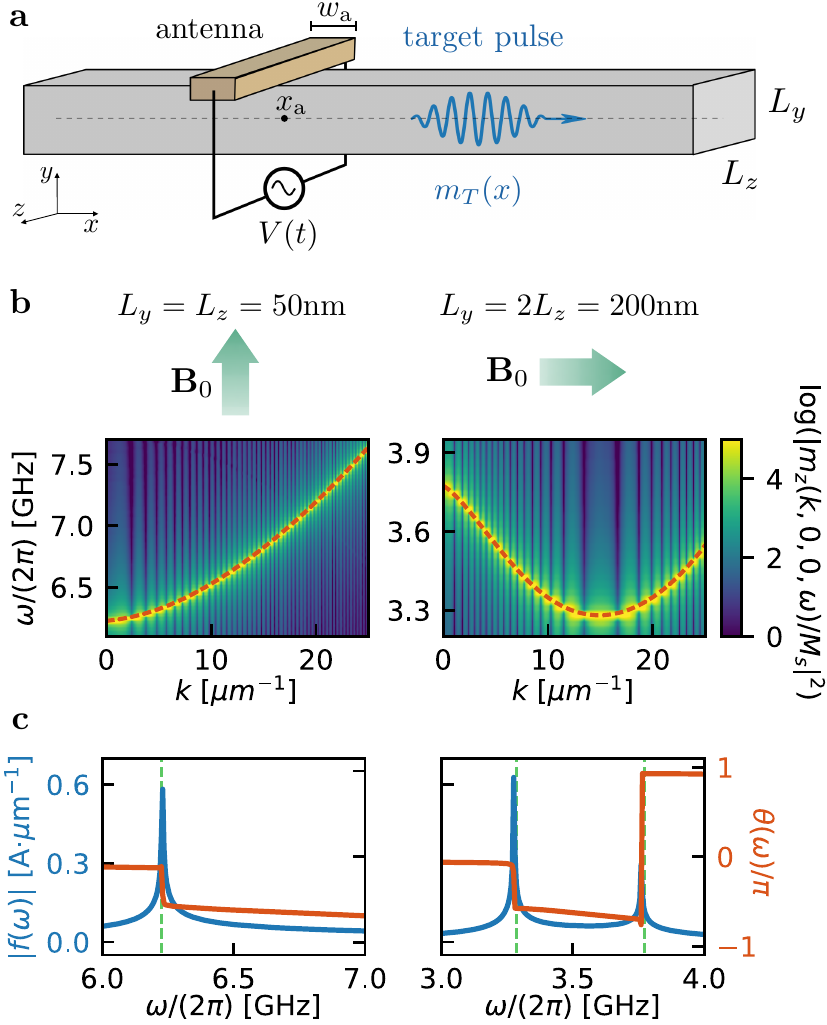}
\par\end{centering}
\caption{\textbf{System charaterization.}
\textbf{a}, Schematic of the generation of a target spin-wave pulse in a single-band waveguide of arbitrary cross-section. Our IDM provides the antenna driving $V(t)$ needed to generate a pulse with a given shape $m_T(x)$ along the waveguide axis using a microwave antenna of width $w_\text{a}$ centered at $x=x_\text{a}$. \textbf{b}, Examples of dispersion relations for Yttrium-Iron-Garnet (YIG) waveguides with a rectangular cross section of $50\text{nm}\times50\text{nm}$ and a transverse bias field (left) and a rectangular cross section of $100\text{nm}\times200\text{nm}$ and a longitudinal bias field (right).  \textbf{c}, Modulus (blue) and phase (orange) of the waveguide transfer function (definition in the main text)  at $\mathbf{r}=\mathbf{0}$ for the same two waveguide configurations, for antenna widths $w_\text{a}=70\text{nm}$ (left) and $w_\text{a}=230\text{nm}$ (right). Here and in the remaining figures we fix $x_\text{a}=0$.  The dashed lines mark the cutoff frequencies at which $d\omega/dk=0$. See Methods section for details on the simulations. \label{fig:fig1}}
\end{figure}

\section*{Method for inverse design of spin-wave pulses}

Pulse engineering comprises the generation of tailored wave packets able to evolve into a desired shape after propagation within a given nanostructure. It provides a method to control wave propagation without nanostructure engineering.
Pulse engineering is used to control optical excitations~\cite{10.1038/s41467-017-01870-0,10.1038/nphoton.2008.249,Roos_2008}, acoustic waves~\cite{Li:19}, or microwaves~\cite{Sharafiev2021visualizingemission} for applications such as quantum information processing~\cite{PhysRevLett.97.050505,PhysRevLett.123.260503}. 
Here, we propose a method for spin-wave pulse engineering based on inverse design (IDM), i.e., a method to determine the  magnetic driving needed to generate a chosen target pulse.

To model spin-wave pulse engineering, we focus on the experimentally relevant system shown in Fig.~\ref{fig:fig1}a, namely an infinite single-band magnonic waveguide with arbitrary cross section, oriented along the $x-$axis and driven by a microwave antenna of width $w_\text{a}$ centered at $x=x_\text{a}$. The antenna generates a magnetic field $\mathbf{B}_1(\mathbf{r},t)=\mathbf{B}_\text{a}(\mathbf{r})V(t)$, with a spatial profile $\mathbf{B}_\text{a}(\mathbf{r})$ given by the antenna geometry and a dimensionless driving $V(t)$ proportional to the applied voltage. This magnetic field generates spin waves, i.e., a propagating dynamic magnetization $\mathbf{m}(\mathbf{r},t)$ on top of the stationary waveguide magnetization. The purpose of the IDM is to determine the driving $V(t)$ needed to generate a spin-wave pulse whose magnetization, at a chosen time $t_f$, transverse position $(y_0,z_0)$, and waveguide arm $x>x_\text{a}$,  has a chosen target pulse shape $m_T(x)$, that is, $\mathbf{e}_0\cdot \mathbf {m}(x,y_0,z_0,t_f)\propto m_T(x)$, where $\mathbf{e}_0$ is an arbitrary unit vector.

The IDM consists of three steps (see Methods section for details): (i) The characterization of the system by calculating the waveguide static magnetization, its dispersion relation $\omega(k)$, and the antenna transfer function $\mathbf{f}(\mathbf{r},\omega)$. The transfer function is defined in frequency domain as the relation between the driving applied to the antenna and the magnetization generated by it at position $\mathbf{r}$, i.e., $\mathbf{m}(\mathbf{r},\omega)\equiv\mathbf{f}(\mathbf{r},\omega)V(\omega)$. Each of these quantities can be calculated efficiently with simple micromagnetic simulations. (ii) The calculation of the \textit{time-dependent} magnetization $\mathbf{m}_T(x_\text{a},y_0,z_0,t)$, at the chosen transverse position ($y_0,z_0$) and at the antenna longitudinal position $x=x_\text{a}$,  which would evolve into the target pulse after free propagation in the waveguide. This magnetization is computed by evolving the pulse \textit{backward} in time while recording the magnetization at $\mathbf{r}=(x_\text{a},y_0,z_0)$. The backward evolution is performed until the whole pulse lies at the other side of the antenna ($x<x_\text{a}$), a time which we set as $t=0$. We use the following approximate expression for the time evolution, valid in the linear regime and for low propagation losses,
\be 
\mathbf{m}_T(\mathbf{r},t)\simeq\frac{1}{\sqrt{2\pi}} \int_\mathbb{R} \text{d}k\left(\mathbf{c}(k;y,z)e^{i(kx-\omega(k)t)}+\text{c.c.}\right), \label{decomposition-pulse}
\ee
where the coefficients $\mathbf{c}(k;y,z)$ are determined by the constraint $\mathbf{e}_0\cdot \mathbf {m}(x,y_0,z_0,t_f)=m_T(x)$. (iii) The computation of the required driving as $V(t) = \mathcal{F}^{-1}\{\mathcal{F}\{\mathbf{e}_0\cdot\mathbf{m}_T(x_\text{a},y_0,z_0,t)\}/[\mathbf{e}_0\cdot\mathbf{f}(x_\text{a},y_0,z_0,\omega)]\}$, where $\mathcal{F}$ and $\mathcal{F}^{-1}$ denote the Fourier and inverse Fourier transforms, respectively. 

\begin{figure*}[]
\noindent \begin{centering}
\includegraphics[scale=1]{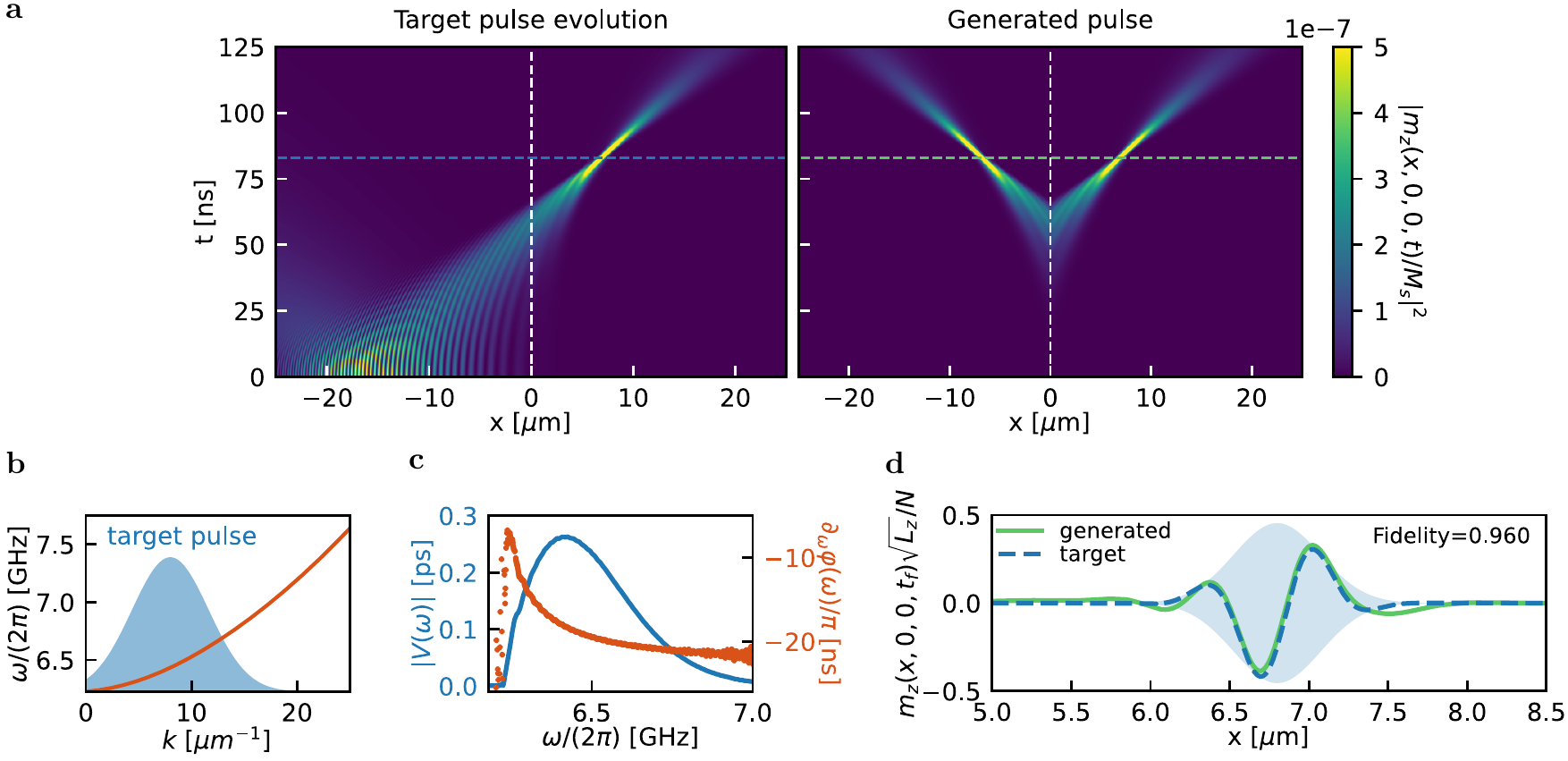}
\par\end{centering}
\caption{ \textbf{Generation of a self-compressing spin-wave pulse in the exchange regime. } \textbf{a}, Magnetization as a function of position along the waveguide and time. Left: Backward evolution of the target pulse \eqnref{target-pulse} in a square YIG waveguide of $L_y=L_z=50$nm width under a transverse bias field $\mathbf{B}_0=(270\text{mT})\mathbf{e}_y$. The pulse parameters are $x_f=6.8\mu$m, $\sigma_f=194\text{nm}$, $k_0=8\mu\text{m}^{-1}$ and $A/M_s=0.004$, where $M_s$ is the saturation magnetization. Right: micromagnetic simulation of the magnetization dynamics under the antenna driving $V(t)$ obtained with our IDM, with chosen parameters $t_f=83$ns, $x_\text{a}=y_0=z_0=0$, and $\mathbf{e}_0 = \mathbf{e}_z$. In both panels, horizontal and vertical lines indicate $t_f$ and the antenna position $x_\text{a}=0$, respectively.
\textbf{b}, Polynomial fit of the waveguide dispersion relation (orange) and envelope of the target pulse \eqnref{target-pulse} in reciprocal space (shaded area).
\textbf{c}, modulus (blue) and derivative of the argument (orange) of the frequency-domain driving function $V(\omega)=\vert V(\omega)\vert e^{i\varphi(\omega)}$ extracted from our IDM.
\textbf{d}, Comparison between the target magnetization (dashed blue line) and the magnetization generated at $t=t_f$ by applying our IDM according to micromagnetic simulations (solid green line). Both magnetizations are square-normalized using $N^2\equiv\int_\mathbb{R}\text{d}x(\mathbf{e}_0\cdot \mathbf{m}(x,y_0,z_0,t_f))^2$. The shaded area shows the envelope of the target pulse.
\label{fig:fig2}}
\end{figure*}

This IDM is valid for any dispersion relation and any target shape not forbidden by physical constraints (e.g., too wide antennas, see example below). It thus provides a universal recipe for the generation of spin-wave pulses in the linear regime. The method is also computationally efficient as steps (ii) and (iii) do not require additional micromagnetic simulations. An essential advantage of this IDM is the use of the approximate expression \eqnref{decomposition-pulse}, which allows to backward-evolve the pulse with minimal computational overhead.

\section*{Relevant examples: self-compressing and rectangular pulses}
	
We demonstrate the performance and universality of our IDM by theoretically studying the generation of pulses relevant for magnonic information processing, and benchmarking it against full micromagnetic simulations. Specifically, we consider the generation of two classes of pulses, namely self-compressing and rectangular pulses, in different waveguides showing different spin-wave regimes (exchange and dipolar). Self-compressing pulses are a family of chirped pulses that compress as they propagate along the waveguide due to the curvature of the dispersion relation~\cite{PhysRevLett.126.103602}. At the time of maximum compression, these pulses have a Gaussian intensity profile whose spot size can be sub-wavelength. In contrast to related ideas, such as non-linear spin-wave bullets  or solitons~\cite{PhysRevLett.81.3769,PhysRevLett.92.117203,PhysRevLett.94.167202,doi:10.1063/1.5041426}, these self-compressing pulses remain within the linear regime, and thus require less power and exhibit less dissipation due to magnon-magnon scattering. In magnonics, self-compressing pulses could be used to locally switch nodes coupled to the waveguide (e.g., magnetic nano-islands~\cite{PappNature2021}), or to partially compensate for propagation losses by compressing all the intensity at the position of the detector, thereby enabling the detection of otherwise too weak signals. Furthermore, although in this work we focus on the linear regime, the compression of one or several pulses at the same point in space can create a strong and localized nonlinear response, a feature that could be used as a synapse trigger in magnon-based neuromorphic computing~\cite{TorrejonNature2017}. As a second example we consider the generation of rectangular pulses, which are the basis of digital information processing as they maximize bit rate and information readability.

\begin{figure}[]
\noindent \begin{centering}
\includegraphics[scale=1.0]{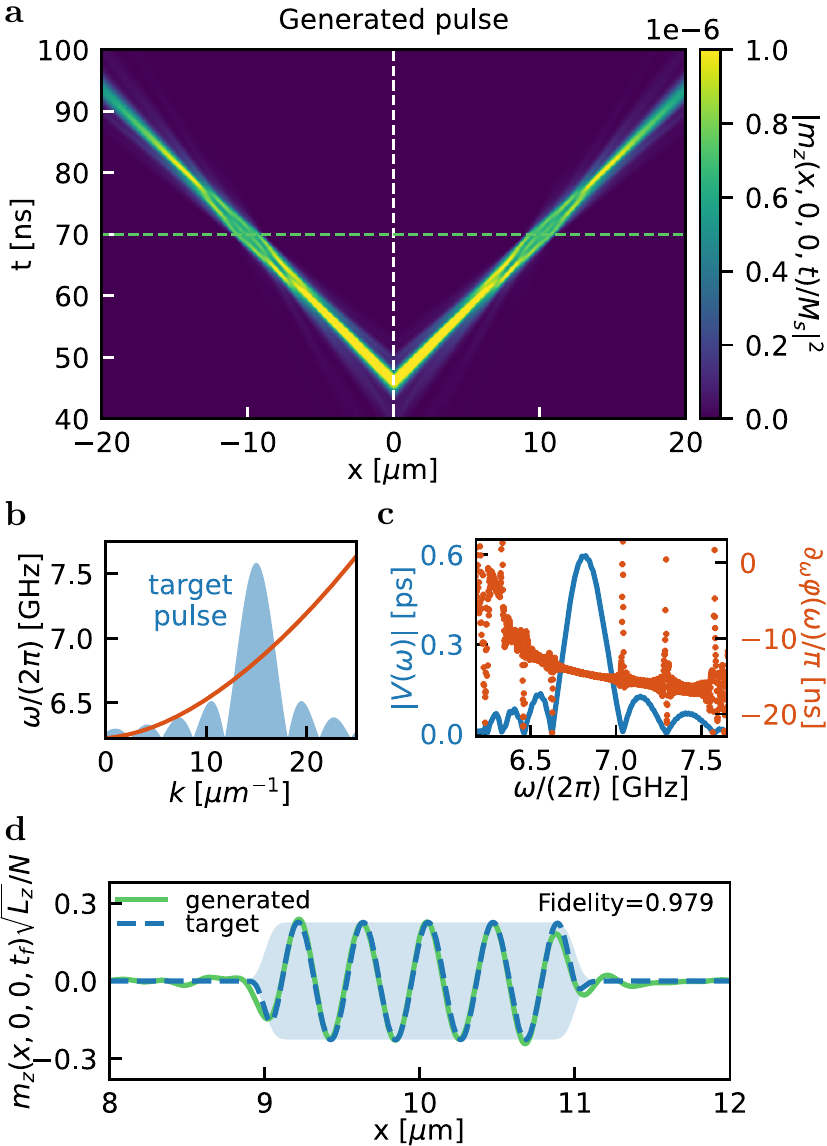}
\par\end{centering}
\caption{\textbf{Generation of a rectangular spin-wave pulse in the exchange regime. } \textbf{a}, Micromagnetic simulation of the magnetization dynamics in a square YIG waveguide of $L_y=L_z=50$nm width,  under a transverse bias field $\mathbf{B}_0=(270\text{mT})\mathbf{e}_y$ and under the antenna driving $V(t)$ given by our IDM. We choose an antenna width $w_\text{a}=70$nm, the target pulse \eqnref{target-square-pulse}, and parameters 
$k_0=15\mu\text{m}^{-1}$, $x_f=10\mu\text{m}$, $\Delta x=2\mu\text{m}$, $A/M_s=0.004$, $\sigma_c=0.05\mu\text{m}$,
$t_f=70$ns, $x_\text{a}=y_0=z_0=0$, and $\mathbf{e}_0 = \mathbf{e}_z$. 
\textbf{b}, Polynomial fit of the dispersion relation (orange) and envelope of the target pulse in reciprocal space (shaded area).
\textbf{c}, Extracted modulus (blue) and derivative of the argument (orange) of the required driving function $V(\omega)=\vert V(\omega)\vert e^{i\varphi(\omega)}$ in frequency domain.
\textbf{d}, Target magnetization (dashed blue line, envelope shown by shaded area) and magnetization generated by our IDM at $t=t_f$ extracted from the micromagnetic simulation (solid green line). Both magnetizations are square-normalized. \label{fig:fig3}}
\end{figure}

First, we focus on the generation of a self-compressing pulse in the exchange regime. We consider a YIG nanowaveguide with a square cross section of $50$nm width and a transverse bias field~\cite{doi:10.1063/5.0045570}, whose dispersion relation and transfer function are shown in Fig.~\ref{fig:fig1}b-c (left panel).  
We use the following Gaussian target pulse~\cite{PhysRevLett.126.103602},
\begin{equation}
m_T(x) = A \exp\left(-\frac{\left(x-x_f\right)^2}{4\sigma_f^2}\right)\cos\left(k_0 x\right),\label{target-pulse}
\end{equation}
where $A$ is the amplitude of the pulse (chosen to ensure that nonlinearity is negligible), $k_0$ is its carrier wavenumber, $x_f$ the point of maximum compression and $\sigma_f$ the width at $x=x_f$. 
The self-compressing behavior of this pulse is evident in its backward time-evolution, shown in Fig.~\ref{fig:fig2}a (left panel). For the parameters used in the figure, the initial pulse width (i.e., standard deviation of intensity, see Methods for details), $\sigma(0)=5.27\mu$m, shrinks to a final value $\sigma(t_f)\approx\sqrt{2}\sigma_f = 295$nm at $t=t_f$ (horizontal dashed line), thus  reaching sub-wavelength spin-wave compression ($\sigma(t_f)k_0/(2\pi) < 1$). 
To generate this pulse using our IDM we consider a narrow antenna of width $w_\text{a}=70$nm, in order to efficiently excite all the pulse wavenumbers $k$ (see Fig.~\ref{fig:fig2}b). Using the backward time-evolution (Fig.~\ref{fig:fig2}a) and the transfer function of this antenna (Fig.~\ref{fig:fig1}c) we obtain the driving $V(t)$ needed to generate the pulse, shown in frequency domain in Fig.~\ref{fig:fig2}c. We then test our result by applying the obtained driving to the antenna and calculating the exact magnetization dynamics using a full micromagnetic simulation (see Methods for details). The resulting generated magnetization profile, which we label $\mathbf{m}_g(\mathbf{r},t)$, is shown in Fig.~\ref{fig:fig2}a (right panel). Note that, although we focus on the right arm of the waveguide ($x>0$), identical pulses are generated at both sides of the antenna as the system is mirror-symmetric around $x=x_a=0$. The  target and generated pulses at $t=t_f$ (both square-normalized) are shown in Fig.~\ref{fig:fig2}d.  To quantify the performance of our method we define the pulse generation fidelity as 
\be\label{fidelity}
F(t)\equiv\frac{\left[\int_{0}^\infty\text{d}x m_g(x,t)m_T(x)\right]^2}{
	\int_{0}^\infty\text{d}x m_g^2(x,t) \int_{0}^\infty\text{d}x m_T^2(x)},
\ee
where $m_g(x,t)\equiv\mathbf{e}_0\cdot \mathbf{m}_g(x,y_0,z_0,t)$.
For the chosen parameters a maximum fidelity $F = 0.963$ is achieved at $t=83.07\text{ns}\approx t_f$, certifying the success of our generation method.  We attribute the small $\approx 4\%$ errors to frequency-dependent propagation losses not considered in \eqnref{decomposition-pulse}.

As a second example we focus on the generation of a rectangular pulse in the exchange regime. We consider the same system as above, namely a $50\text{nm}\times50$nm square YIG waveguide excited by an antenna with $w_a=70$nm. We use the following rectangular target pulse,
\be
m_T(x)=\frac{A}{2} \left[\text{erf}(x_+)+\text{erf}(x_-)\right]\cos(k_0 x),\label{target-square-pulse}
\ee
where ${x_\pm\equiv (\Delta x \pm 2 (x-x_f) )/(2\sqrt{2}\sigma_c)}$ and $\text{erf}(x)$ is the error function. Here, $k_0$ is the carrier wavenumber, $A$ the amplitude, $x_f$ the center of the pulse, $\Delta x$ its spatial extension and $\sigma_c$ determines the curvature at the edge of the rectangular envelope. As shown in Fig.~\ref{fig:fig3}, this pulse can be successfully generated using the IDM method detailed above. Specifically, short rectangular pulses of duration $\approx 5$ns can be generated with fidelity $0.98$.

\begin{figure}[]
\noindent \begin{centering}
\includegraphics[scale=1]{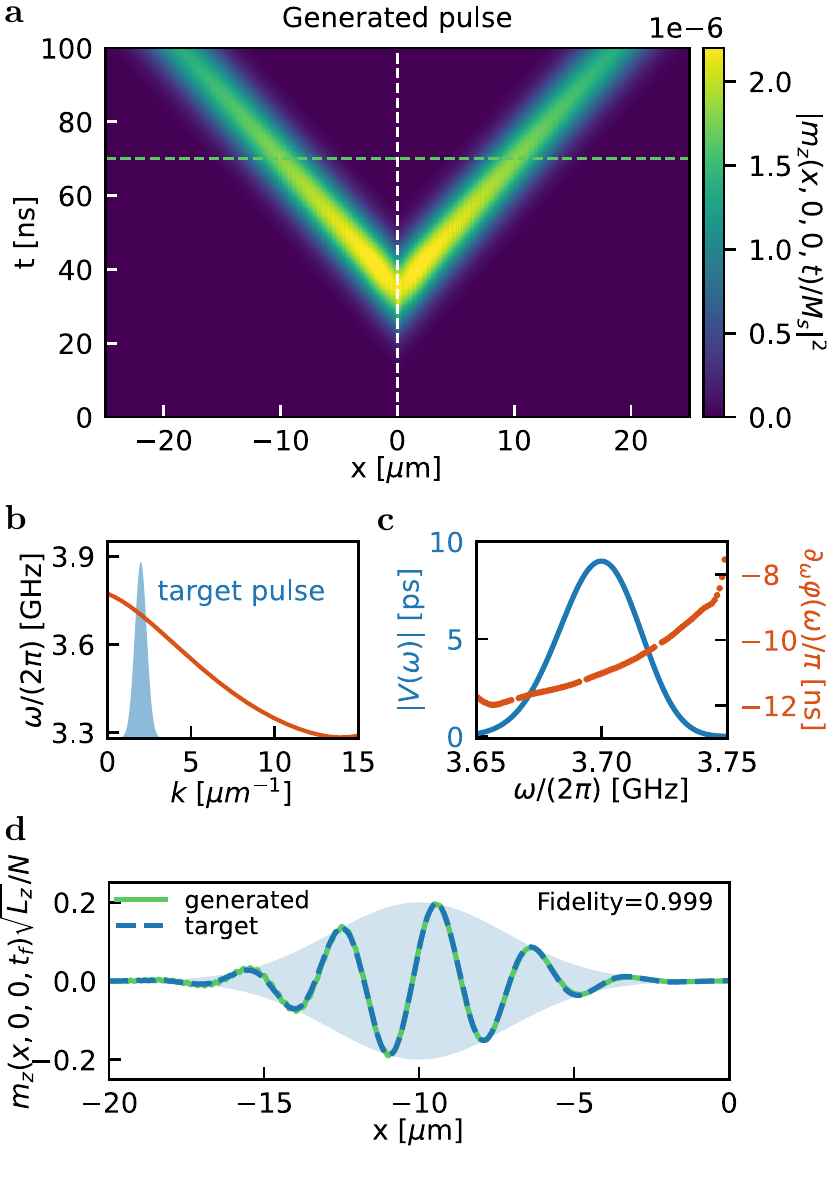}
\par\end{centering}
	\caption{\textbf{Generation of a weakly self-compressing pulse in the dipolar regime. } \textbf{a}, Micromagnetic simulation of the magnetization dynamics in a YIG rectangular waveguide of $200\text{nm}\times100$nm cross-section  under a longitudinal bias field $\mathbf{B}_0=(50\text{mT})\mathbf{e}_x$ and the antenna driving $V(t)$ given by our IDM. We choose an antenna width $w_\text{a}=230$nm and parameters $x_f=-10\mu$m, $\sigma_f=2\mu$m, $k_0=2\mu\text{m}^{-1}$,  $A/M_s=0.004$, $t_f=70$ns, $x_\text{a}=y_0=z_0=0$, and $\mathbf{e}_0 = \mathbf{e}_z$. 
	\textbf{b}, Polynomial fit to the dispersion relation of the waveguide (orange) and target pulse envelope in reciprocal space (shaded area).
	\textbf{c}, Extracted modulus (blue) and derivative of the phase (orange) of the required driving function $V(\omega)=\vert V(\omega)\vert e^{i\varphi(\omega)}$ in frequency domain. \textbf{d}, Target magnetization (dashed blue, envelope shown by shaded area) and magnetization generated at $t=t_f$ extracted from the micromagnetic simulation (solid green). Both magnetizations are square-normalized.
	\label{fig:chirped-pulse-BVW}}
\end{figure}

As a third example, in Fig.~\ref{fig:chirped-pulse-BVW} we demonstrate the generation of a weakly self-compressing pulse in the backward-volume wave regime of a YIG waveguide with a rectangular cross section of $200\text{nm}\times100$nm. In this setup, as the derivative of the dispersion is negative (see right panel of Fig.~\ref{fig:fig1}b), the target pulse is chosen at the opposite waveguide arm $x<x_a$ and leftward-propagating.
Moreover, for this waveguide the dispersion relation is degenerate in the range $3.30\text{ GHz}\lesssim \omega/(2\pi)\lesssim 3.75\text{ GHz}$, i.e., there are two spin wave modes with the same energy and different wavenumbers. To guarantee that the spin wave pulse is only generated in the region of negative derivative of the dispersion we choose a larger antenna ($w_a=230$nm), which is unable to excite wavenumbers larger than $\approx \pi/w_a = 13.6 \mu \text{m}^{-1}$. The resulting pulse generation has a fidelity of $0.999$. The different classes of pulses considered in this work show that our method is universal and enables near-perfect generation of spin-wave pulses of arbitrary shape in both the exchange and the dipolar regime. 

\begin{figure}[]
\noindent \begin{centering}
\includegraphics[scale=1]{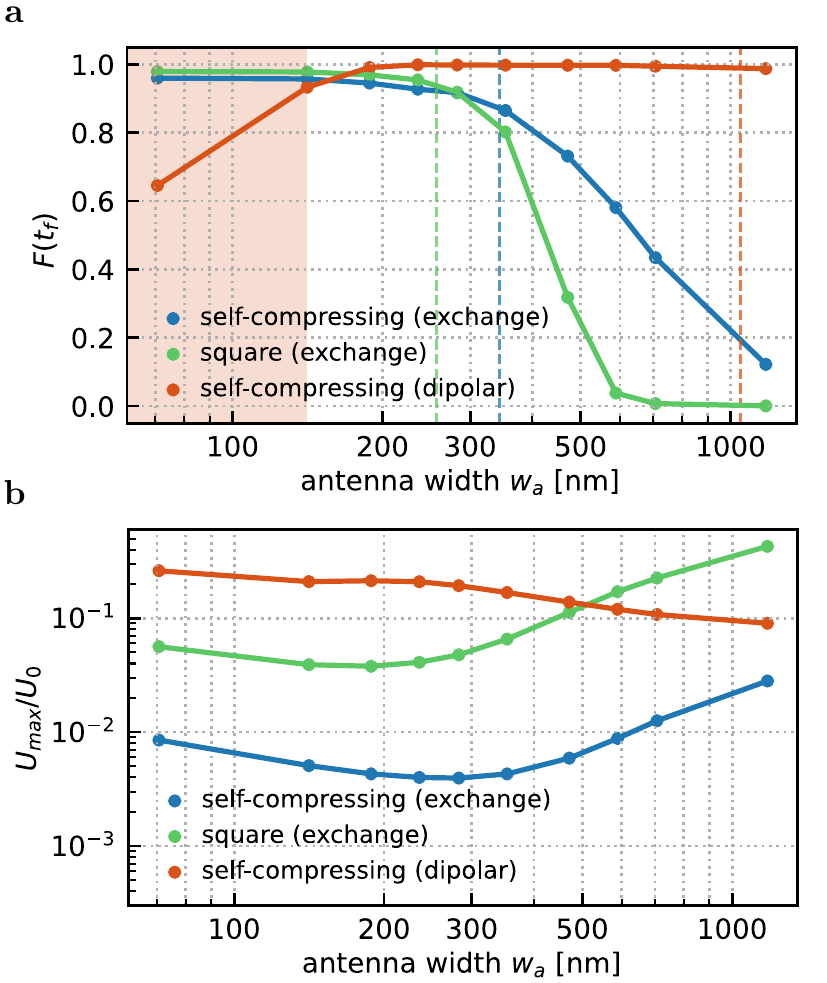}
\par\end{centering}
\caption{ \textbf{Pulse generation using wide antennas}. $\mathbf{a}$ Pulse generation fidelity  as a function of antenna width $w_\text{a}$ for the three pulses shown in Figs.~\ref{fig:fig2}-\ref{fig:chirped-pulse-BVW}. 
For each antenna width the pulse has been rescaled so that its peak amplitude remains constant.
The dashed vertical lines indicate the minimum wavelengths $\lambda_\text{min}$ of each pulse ($\lambda_\text{min}/2$ for the dipolar-regime pulse). The shaded area indicates the region where the fidelity for the dipolar-regime pulse decreases below $0.9$ due to the excitation of degenerate modes. $\mathbf{b}$, Maximum value $U_\text{max}$ of the instantaneous magnetic energy of the driving magnetic field as a function of antenna width $w_\text{a}$, normalized to the energy associated to the homogeneous field $\mathbf{B}_0$ used in each case. See Methods for the definitions of minimum wavelength  and energy. In both panels, solid lines are a guide to the eye.  \label{fig:fidelity}}
\end{figure}

Although the narrow antennas used in Figs.~\ref{fig:fig2} and \ref{fig:fig3} are experimentally feasible, wider antennas are in practice desirable as they are simpler to fabricate and provide higher spin-wave excitation efficiency. In Fig.~\ref{fig:fidelity}a we study the pulse generation fidelity as a function of antenna width $w_\text{a}$, for the three example pulses shown above. As an antenna cannot generate spin waves with wavelengths smaller than $\approx 2 w_\text{a}$, the generation fidelities are bound to decrease for $w_\text{a} \gtrsim \lambda_{\rm min}$, with $\lambda_{\rm min}$ the minimum wavelength of the pulse (see Methods section for a definition). Generation fidelities above $90\%$ can still be achieved at $w_\text{a}\lesssim 300$nm for the pulses of Figs.~\ref{fig:fig2} and \ref{fig:fig3}, and beyond $w_\text{a}= 1\mu$m for the spatially much wider backward-volume pulse of Fig.~\ref{fig:chirped-pulse-BVW}. 	
For the latter, the decrease in fidelity for narrow antennas, indicated by the shaded area, stems from the excitation of unwanted, high-wavenumber modes in the degenerate dispersion relation (Fig.~\ref{fig:fig1}b, right panel). Within the regions of high fidelity, the energy required to generate the pulse is reduced for wider antennas. 
This is indicated in Fig.~\ref{fig:fidelity}b where, as a figure of merit for energy cost, we display the maximum value of the instantaneous energy stored in the waveguide by the antenna driving field (see details in Methods). We emphasize that the low fidelity regions in Fig.~\ref{fig:fidelity}a do not manifest a failure of our method but an unphysical choice of the target pulses, which cannot be generated by antennas of certain widths.

\pagebreak

\section*{Conclusion} \label{outlook}

We have proposed a method for universal spin-wave pulse engineering based on inverse design. The method provides, in a numerically efficient way, the time-dependent driving which can be applied to a narrow antenna to generate an arbitrary target spin-wave pulse in the linear regime. Our concept is universal as it applies to arbitrary waveguide and antenna geometries, and to both the exchange and dipolar regimes. Using micromagnetic simulations, we have theoretically shown high-fidelity generation of relevant pulses for magnonics. Specifically, we have predicted the generation of a self-compressing pulse which compresses into a sub-wavelength spot of width $295$nm, with fidelity 0.96. At the compression spot the pulse intensity is 3.7 times higher than the peak intensity of the pulse right after the driving even in the presence of damping. Moreover, we have predicted the generation of rectangular pulses of $\sim$5ns duration with fidelity 0.98. Both pulses can be generated in the exchange regime with antennas as wide as 300nm with fidelity $>0.9$. Finally, we have shown how even wider antennas of widths $\sim 1\mu$m can be used to generate a weakly self-compressing pulse in the dipolar regime with a fidelity of 0.999. 
The presented inverse design method can be extended to other magnonic structures, and could be refined at the cost of higher computational complexity, e.g., by including wavelength-dependent spin wave losses in the backward propagation step. 
Our results could enable fast magnon-based information processing at the nanoscale and pave the way to implementing further nanophotonics-inspired strategies in magnonics to devise magnon-based quantum technological platforms.

\bibliography{Mybib.bib}

\clearpage
 
\section*{Methods}
	
\subsection{Waveguide and antenna characterization}
	
We determine the waveguide static magnetization and dispersion relation using MuMax3~\cite{mumax3}. To compute the dispersion relation we evolve the magnetization under a magnetic field $\mathbf{B}_1(\mathbf{r},t)\propto\mathbf{B}_\text{a}(\mathbf{r})\text{sin}\left(\omega_\text{max}t\right)/t$, able to excite spin waves in a wide range of frequencies $(\omega<\omega_{\rm max})$ and wavevectors. 
Applying a two-dimensional Fourier transform to the simulated magnetization leads to the excitation spectrum of the spin waves, 
\be
\mathbf{m}(k,0,0,\omega)=\frac{1}{2\pi}\int_\mathbb{R}\int_\mathbb{R}\text{d}x\,\text{d}t\,\mathbf{m}(x,0,0,t)e^{-i(kx+\omega t)}.\label{m_spectrum}
\ee
Two examples of these spectra for different waveguides and $\omega_{\rm max}/(2\pi) = 10$GHz are shown in Fig.~\ref{fig:fig1}b. The spin wave frequency for each wavenumber $k$ can then be extracted as the maximum of the excitation spectrum for such a wavenumber (red dashed curves in the figure).
	
We model the field generated by the antenna at the waveguide by the Gaussian function
\be
\mathbf{B}_\text{a}(\mathbf{r})=\frac{|\mathbf{B}_0| L_z}{\sqrt{2\pi} \sigma_\text{a}}\exp\left( -\frac{x^2}{2\sigma_\text{a}^2}\right)\mathbf{e}_z.\label{gaussian-30nm}
\ee
The physical width of the antenna can be identified with the full width at half maximum of the field profile,  $w_\text{a}=2\sqrt{2\log(2)}\sigma_\text{a}\approx 2.3\sigma_\text{a}$.
To determine the antenna transfer function, we first define the Fourier transform for a vector function as $\mathbf{v}(t)$ as $\mathcal{F}\{\mathbf{v}(t)\} \equiv (\sqrt{2\pi})^{-1}\int_\mathbb{R} \text{d}t \mathbf{v}(t)\exp({-i\omega t})$. Then, for every chosen antenna (i.e., for every value of $\sigma_\text{\rm a}$) we perform one micromagnetic simulation of the  magnetization dynamics in the presence of the driving field \eqnref{gaussian-30nm}, using an impulse test driving $V_{\rm test}(t)=\delta(t)$. For this driving,  the transfer function in time domain is simply proportional to the magnetization field, $\mathbf{f}(\mathbf{r},t)=\sqrt{2\pi}\mathbf{m}(\mathbf{r},t)$. To determine the transfer function in Fig.~\ref{fig:fig1}c (left panel) we use an impulse driving $V(t)=0.15$ during a single time step of 1ps and $V(t)=0$ afterward. The antenna field width is chosen as $\sigma_\text{\rm a}=30$nm, corresponding to an antenna width of about  $w_\text{a}\approx 70$nm. In Fig.~\ref{fig:fig1}c (right panel) we choose $\sigma_\text{\rm a}=100$nm, corresponding to  $w_\text{a}\approx230$nm, and an amplitude of the impulse driving $V(t)=0.4$ during a single time step of 1ps.

In Figs.~\ref{fig:fig2} and \ref{fig:fig3} we model the system as a finite waveguide of dimensions  $(80\mu\text{m}\times50\text{nm}\times50\text{nm})$, with material parameters for Yttrium-Iron-Garnet (YIG) ~\cite{doi:10.1063/5.0045570}, namely saturation magnetization $M_s=140.7\text{kA}\cdot\text{m}^{-1}$, exchange constant $A_\text{ex}=4.2\text{pJ}\cdot\text{m}^{-1}$, Gilbert damping parameter $\alpha=1.75\cdot10^{-4}$, and a cell size of $(10\times6.25\times6.25)\text{nm}^3$.  For Fig.~\ref{fig:chirped-pulse-BVW} we use a waveguide of dimensions  $(80\mu\text{m}\times200\text{nm}\times100\text{nm})$ with the same material parameters described above, and a cell size of $(10\times10\times10)\text{nm}^3$.

\subsection{Backward propagation of the pulse}

In order to perform the backward propagation in a fast way, we obtain an analytical approximation for the dispersion relation $\omega(k)$ by fitting the maxima of the magnetization spectrum to a sixth-order polynomial in $k$. The resulting polynomial approximations, shown in panel b of  Figs.~\ref{fig:fig2} to \ref{fig:chirped-pulse-BVW}, are then used to integrate \eqnref{decomposition-pulse} numerically. The maximum error incurred by this polynomial approximation is below $0.25\%$ for all the wavenumbers in Figs.~\ref{fig:fig2} and \ref{fig:fig3} and below $0.45\%$ in Fig.~\ref{fig:chirped-pulse-BVW}. The integral is computed back to a time $t=0$ defined as the time at which $99.5\%$ of the pulse lies at the opposite side of the antenna, i.e., 
\be
\left|\frac{\int_{L_0}^{x_\text{a}} \text{d}x (\mathbf{e}_0\cdot \mathbf{m}_T(x,y_0,z_0,0))^2}{\int_{\mathbb{R}}\text{d}x (\mathbf{e}_0\cdot \mathbf{m}_T(x,y_0,z_0,0))^2}\right|= 0.995,
\ee
where $L_0\equiv-\infty$ ($L_0\equiv+\infty$) for  rightward (left)-propagating pulses.

\subsection{Definition of pulse width and minimum wavelength}
	
We define the time-dependent width of a given pulse as the standard deviation of the pulse position, i.e.,
\be
\sigma(t)\equiv\sqrt{\int_\mathbb{R}\text{d}x\frac{(\mathbf{e_0}\cdot \mathbf{m}(x,y_0,z_0,t))^2}{N^2(t)}(x-\bar{x}(t))^2}.\label{width}
\ee
where $N^2(t)\equiv\int_\mathbb{R}\text{d}x (\mathbf{e}_0\cdot \mathbf{m}(x,y_0,z_0,t))^2$ is a normalization factor and $\bar{x}(t)\equiv\int_\mathbb{R}\text{d}x \,x \,(\mathbf{e}_0\cdot \mathbf{m}(x,y_0,z_0,t))^2/N^2(t)$ is the  mean value of the pulse position.

The minimum  wavelength of each pulse is defined as $\lambda_{\rm min}\equiv2\pi/k_{\rm max}$, where we define $k_\text{max}$ such that 
\be
\frac{\int_{-\infty}^{k_\text{max}}\text{d}k |\mathcal{F}^{-1}\{m_T(x)\}|}{\int_{-\infty}^\infty\text{d}k |\mathcal{F}^{-1}\{m_T(x)\}|}=0.997.\label{kmax}
\ee
Here $m_T(k) \equiv (\sqrt{2\pi})^{-1}\int_{\mathbb{R}}\text{d}t m_T(x) \exp(ikx) $ is the inverse Fourier transform of the target pulse $m_T(x)$ in position space. For the Gaussian pulse \eqnref{target-pulse}, the condition  \eqnref{kmax} corresponds to $k_{\rm max} = k_0 + 2/\sigma_f$.
	
\subsection{Energy cost of generating the driving field}

The energy required to generate a pulse is lower-bounded by the total energy required to generate the driving field. We define the following figure of merit for the latter,
\begin{equation}\label{Umax}
    U_{\rm max} \equiv \max_t\int_{V} \text{d}^3\mathbf{r}\int_{-\infty}^t\text{d}
    t' \textbf{H}_1(\mathbf{r},t')\cdot\frac{\partial}{\partial t'} \textbf{B}_1(\mathbf{r},t'),
\end{equation}
where $V$ denotes the volume of the waveguide. Equation (\ref{Umax})  corresponds to the maximum value of the instantaneous magnetic energy held inside the waveguide due to the presence of a driving field $\textbf{B}_1(\mathbf{r},t)$~\cite{StancilBook2009}. Both fields in the integrand of \eqnref{Umax} are assumed to vanish at $t'=-\infty$ and are related in frequency domain by
\begin{equation}
   \textbf{H}_1(\mathbf{r},\omega) = \frac{1}{\mu_0}\bar{\boldsymbol{\mu}}^{-1}(\omega)\textbf{B}_1(\mathbf{r},\omega),
\end{equation}
where $\mu_0$ is the vacuum permeability and $\bar{\boldsymbol{\mu}}(\omega)$ is the relative permeability tensor.

In frequency domain, all pulses considered in this article have central frequencies near $\omega_0\equiv |\gamma\mathbf{B}_0|$, where $\mathbf{B}_0$ is the homogeneous bias field and $\gamma$ is the gyromagnetic ratio, and widths much smaller than $\omega_0$. We can thus approximate $\bar{\boldsymbol{\mu}}^{-1}(\omega) \approx \bar{\boldsymbol{\mu}}^{-1}(\omega_0)$ in the above expression. Using the identity $\mathbf{B}_1(\mathbf{r},t)=\mathbf{B}_\text{a}(\mathbf{r})V(t)$ and assuming that the only non-zero component of $\mathbf{B}_\text{a}(\mathbf{r})$ is oriented along the unit vector $\mathbf{e}_z$, we cast the energy as 
\begin{equation}
    U_{\rm max} = \frac{L_y L_z}{2\mu_0}\left[\bar{\boldsymbol{\mu}}^{-1}(\omega_0)\right]_{zz}\int_\mathbb{R}\text{d}x\left|\mathbf{B}_\text{a}(\mathbf{r})\right|^2\max_tV^2(t).
\end{equation}
The energy $U_\text{max}$ as a function of the antenna width is displayed in Fig.~\ref{fig:fidelity}b. It is normalized to a reference energy $U_0$, namely the energy stored by the constant homogeneous bias field $\mathbf{B}_0$ in a section of the waveguide large enough to contain the pulse at all times, given by $U_0=L_x L_y L_z |\mathbf{B}_0|^2/(2\mu_0)$ with $L_z=80\mu$m taken as the length of our micromagnetic simulation domain. In particular, $U_0=5.83$fJ  for the homogeneous field used in the generation of pulses in the exchange regime and $U_0=1.59$fJ for the dipolar pulse. To compute the value of  $U_\text{max}$, we approximate the inverse permeability tensor by its Polder susceptibility expression~\cite{StancilBook2009},
\begin{equation}
    \left[\bar{\boldsymbol{\mu}}^{-1}(\omega_0)\right]_{zz} \approx \frac{1}{2+(\mu_0M_s/|\mathbf{B}_0|)},
\end{equation}
where $M_s$ is the saturation magnetization of the waveguide.

\section*{Acknowledgements}

We acknowledge discussions with O. Dobrovolskiy,  J. J. Garc{\'i}a Ripoll, and T. A.  Gustafsson. This work was supported by the Austrian Science Fund (FWF) through Project No. I 4917-N (MagFunc). SK acknowledges the support by the H2020-MSCA-IF under the grant number 101025758 (OMNI).

\end{document}